\documentclass[twocolumn]{aastex631}

\usepackage{amsmath}
\usepackage[utf8]{inputenc}
\usepackage{xcolor}

\begin{document}

\title{Evaluation of several explanations of the strong X-ray polarization of the black hole X-ray binary 4U\,1630--47}
\author[0000-0002-1084-6507]{Henric Krawczynski}
\affiliation{Physics Department, McDonnell Center for the Space Sciences, and Center for Quantum Leaps, Washington University in St. Louis, St. Louis, MO 63130, USA}
\email{Corresponding authors:
H. Krawczynski (krawcz@wustl.edu), Yajie Yuan
(yajiey@wustl.edu), Alexander Y. Chen (cyuran@wustl.edu).}
\author[0000-0002-0108-4774]{Yajie Yuan}
\affiliation{Physics Department and McDonnell Center for the Space Sciences, Washington University in St. Louis, St. Louis, MO 63130, USA}
\author[0000-0002-4738-1168]{Alexander Y. Chen}
\affiliation{Physics Department and McDonnell Center for the Space Sciences, Washington University in St. Louis, St. Louis, MO 63130, USA}
\author[0000-0002-9705-7948]{Kun Hu}
\affiliation{Physics Department, McDonnell Center for the Space Sciences, and Center for Quantum Leaps, Washington University in St. Louis, St. Louis, MO 63130, USA}
\author[0000-0001-5256-0278]{Nicole Rodriguez Cavero}
\affiliation{Physics Department, McDonnell Center for the Space Sciences, and Center for Quantum Leaps, Washington University in St. Louis, St. Louis, MO 63130, USA}
\author[0009-0002-2488-5272]{Sohee Chun}
\affiliation{Physics Department, McDonnell Center for the Space Sciences, and Center for Quantum Leaps, Washington University in St. Louis, St. Louis, MO 63130, USA}
\author[0000-0002-5250-2710]{Ephraim Gau}
\affiliation{Physics Department, McDonnell Center for the Space Sciences, and Center for Quantum Leaps, Washington University in St. Louis, St. Louis, MO 63130, USA}
\author[0000-0002-5872-6061]{James F. Steiner}
\affiliation{Harvard-Smithsonian Center for Astrophysics, 60 Garden Street, Cambridge, MA 02138, USA}
\author[0000-0003-0079-1239]{Michal Dovčiak}
\affiliation{Astronomical Institute of the Czech Academy of Sciences, Boční II 1401/1, 
14100 Praha 4, Czech Republic}

\begin{abstract}
The {\it Imaging X-ray Polarimetry Explorer (IXPE)}  observations of the X-ray binary 4U\,1630--47 in the high soft state revealed high linear polarization degrees (PDs) rising from 6\% at 2\,keV to 10\% at 8\,keV.  We discuss in this letter three different mechanisms that impact the polarization of the observed X-rays: the reflection of gravitationally lensed emission by the accretion disk, reprocessing of the emission in outflowing plasma, and electron and ion anisotropies in the accretion disk atmosphere. 
We conducted detailed raytracing studies to evaluate the impact of the reflection of strongly gravitationally lensed emission on the PDs. Although the reflected emission can produce high PDs in the high-energy tail of the thermal emission component, we do not find models that describe the PDs and are consistent with independent estimates of the source distance.
We discuss the energetics of another proposed mechanism: the emission or scattering of the X-rays in 
mildly relativistically moving plasma outflows. We argue that these models are disfavored as they require 
large mechanical luminosities on the order of, or even exceeding, the Eddington Luminosity. 
We investigated the impact of electron and ion anisotropies, but find that their impact on the observed PDs are likely negligible.
We conclude with a discussion of all three effects and avenues for future research.  
\end{abstract}


\section{Introduction}
\label{s:intro}
The \textit{Imaging X-ray Polarimetry Explorer (IXPE)} \citep[\textit{IXPE},][]{ixpe} launched on Dec. 9, 2021
measured or constrained the polarization 
of the X-rays from several Black Hole X-ray 
Binaries (BHXRBs).
In the thermally dominated soft state, {\it IXPE} observations allow us to test the standard geometrically thin, optically thick accretion disk model, and - if verified - to constrain the source inclination \citep{2009ApJ...691..847L} and black hole spin \citep{2009ApJ...701.1175S}. In the hard state, the {\it IXPE} observations constrain the geometry, location, and physical properties the hard X-ray emitting corona and the inclination of the accretion flow \citep{2022Sci...378..650K}.  
To date, {\it IXPE} has observed six X-ray BHXRBs with data in the public domain \citep{2024mbhe.confE..39C}.  Three black holes were observed in more than one emission state: Cyg X-1 in the hard state (HS) and in the High Soft State (HSS) \citep{2022Sci...378..650K,2023ATel16084....1D,2024MNRAS.52710837J},  4U~1630$-$47 in the HSS and in the Steep Power Law (SPL) state \citep{2024ApJ...964...77R,2023ApJ...958L...8R}, and 
Swift J1727.8-1613 in the HS, an intermediate disk-dominated state, and in a dim hard state \citep{2023ApJ...958L..16V,2024ApJ...968...76I,2024A&A...686L..12P}. The black holes LMC X-1 \citep{2023MNRAS.526.5964P}, LMC X-3 \citep{2024ApJ...960....3S} and 4U 1957$+$115 \citep{2024A&A...684A..95M} were observed in the soft state.
The observations of the black hole candidate Cyg X-3 in the hard state led to the reclassification of the source as an Ultraluminous X-ray source (ULX), which emits strongly polarized X-rays owing to the reflection of  X-rays off funnel walls \citep{2024NatAs.tmp..117V}. The source has most recently also been observed in the soft state
\citep[][Rodriguez Cavero et al., in preparation]{veledina2024ultrasoftstatemicroquasarcygnus}.

In this paper, we focus on 4U 1630--47, a low-mass X-ray binary (LMXB) with recurrent outbursts 
every 2--3 years \citep{1998ApJ...494..753K, 2015MNRAS.450.3840C}. 
The {\it IXPE} observations of the HSS are particularly interesting as they probe the polarization of the emission 
from the optically thick geometrically thin accretion disk, and thus allow us to test classical thin disk theory 
\citep{1973A&A....24..337S}. The {\it IXPE} observations 
revealed PDs  increasing from $\sim$6\% at  2~keV to $\sim$10\% at 8~keV in the HSS \citep{2024ApJ...964...77R}
and increasing from $\sim5\%$ at 2~keV  to $\sim$8\% at 8~keV in the SPL state. For both observations, the polarization angles (PAs) did not exhibit statistically significant variations with energy or time.

The PDs measured in the HSS are high compared to theoretical expectations   \citep{2009ApJ...691..847L,2009ApJ...701.1175S,2022ApJ...934....4K,2019ApJ...875..148Z,2020MNRAS.493.4960T,2024ApJ...964...77R}.
According to Chandrasekhar's classical treatment, the emission from an electron scattering atmosphere 
creates PDs of between 0\% ($0^{\circ}$ inclination) and 11.71\% ($90^{\circ}$ inclination) \citep{Chandra:60}. 
The inclination of 4U\,1630--47 is estimated to be $\sim$65$^{\circ}$ \citep{1998ApJ...494..747T,1998ApJ...494..753K} 
for which Chandrasekhar’s PD equals 2.8\%. Furthermore, strong gravitational lensing tends to reduce the observed PDs
as the polarization of photons that traveled through the curved spacetime close to the  black hole and scattered off the accretion disk partially cancels the polarization of the photons emitted  further away from the black hole \citep{2024ApJ...964...77R}. The latter authors manage to explain the {\it IXPE} results with a model combining three effects which increase the PD of the emission: 
a low  black hole spin (reducing the effects from strong lensing),  the emission of highly polarized X-rays 
by a partially ionized plasma, and the emitting plasma streaming with 50\%  of the speed of light away from the accretion disk.  The latter effect increases the observed PDs as X-rays emitted at higher local inclinations
reach the observer at the binary inclination 
owing to relativistic aberration. 
The same authors discuss two other scenarios but find that they cannot explain the {\it IXPE} results:  
slim disks generate higher PDs than thin disks \citep{2023ApJ...957....9W} but still not as high as the observed ones;
the reflection of the X-rays off a wind cannot explain the rise of the PDs with energy \citep{2024NatAs.tmp..117V}.
Detailed modeling of the emission from systems with winds confirm these results and show that the wind reflection
reduces rather than increases the PD of the emission \citep{2024MNRAS.527.7047T}. 

In this paper, we discuss three mechanisms that may contribute to the polarization of the X-rays from 4U\,1630--47. Section \ref{kerrClight} presents the results from fitting the 4U\,1630--47 data with the general relativistic raytracing model {\tt kerrClight}. Although the reflection of strongly gravitationally lensed disk emission increases the observable PDs, we do not find model parameters that fit the high PDs of 4U\,1630--47 and are consistent with independent constraints on the source distance. We discuss alternative models involving outflowing emitting plasmas in Sect.\,\ref{outflowing}. We find that these models require plasma outflows with high mechanical luminosities. Section \ref{anisotropies} discusses  the possible impact of 
electron and/or ion anisotropies on the polarization  of the disk emission. Although order unity anisotropies can lead to highly polarized Bremsstrahlung and Compton scattered emission, we find that this effect is unlikely to play a role in the HSS of BHXRBs as Coulomb collisions limit particle anisotropies to very small levels\footnote{We thank J.\ Poutanen for emphasizing this point in a discussion of an early draft of this paper}. We conclude with a discussion of our findings in Sect.\ \ref{discussion}.

The polarized radiation transport code described in the appendix has been published as a Zenodo archive \citep{https://doi.org/10.5281/zenodo.13904193}.
\section{Fitting of the {\it IXPE}, {\it NICER}, and {\it NuSTAR} HSS data with 
{\tt kerrClight}} \label{kerrClight}
In this section, we analyze the {\it IXPE}, NICER, and {\it NuSTAR} observations of 4U\,1630--47 in HSS using the data sets described in \citep[Appendix A]{2024ApJ...964...77R}. 
The {\it IXPE} observations were acquired between Aug.\,23, 2022 and Sept.\,2, 2022 and were reduced with the standard methods \citep{Baldini2022};  27 ksec of NICER observations were acquired between Aug.\,22, 2022 and Sept.\,1, 2022 and were reduced with the {\tt NICERDAS} software (v. 09) of the HEASOFT web package \citep{2014ascl.soft08004N};  lastly, a total of $\sim~70$ ksec of {\it NuSTAR} observations were acquired between Aug.\,25, 2022 and Aug.\,29, 2022 and were reduced with the {\tt NuSTARDAS} software package (v. 2.1.2), also part of the HEASOFT suite. We use the {\it IXPE} data from 2-8\,keV, the {\it NICER} data from 2-8 keV, and the {\it NuSTAR} data from 3\,keV to 40 keV.

We analyze the data with the {\tt kerrClight} X-ray fitting model. The model is a simplified version of the more general {\tt kerrC} fitting model \citep{2022ApJ...934....4K}, which assumes a geometrically thin, optically thick Novikov-Thorne accretion disk interacting with 3-D coronas of hot plasma of different shapes, i.e.\ wedge-shaped coronas sandwiching the accretion disk, and cone-shaped coronas centered on the black hole spin axis. 
The accretion disk extends from the radial distance of the
Innermost Circular Orbit $r_{\rm ISCO}$ to 100$\,r_{\rm g}$ ($r_{\rm g}\,\equiv\,G\,M\,/\,c^2$ with $G$ being the gravitational constant, $M$ the black hole mass, and $c$ the speed of light), and the code assumes that the angular momentum vectors of the black hole and the accretion disk are aligned. 
The code is based on raytracing photons from the disk to the observer accounting for scatterings off the disk and in the corona. Assuming a high ionization of the photospheric and coronal plasma, {\tt kerrClight} implements the scattering off the disk based on Chandrasekhar’s prescription for the reflection of polarized beams off an indefinitely deep  scattering atmosphere \citep{Chandra:60}.  In contrast to {\tt kerrC}, {\tt kerrClight} does not account for the reprocessing of 
X-rays in a partially ionized photosphere \citep{2010ApJ...718..695G,2013ApJ...768..146G}. 
Indeed, detailed modeling of the 4U\,1630--47 data largely validate this assumption \citep[][Appendix B]{2024ApJ...964...77R}.   
We only consider sandwich coronas here, led by the constraints on the shape of the corona of Cyg X-1 in the hard state  \citep{2022Sci...378..650K}. 
The {\tt kerrClight} model parameters describing the black hole system and the corona are the black hole mass $M$, 
the black hole spin parameter $a$ ($-1 \le a \le 1$), inclination (angle of the observer relative to the black hole spin axis) $i$, black hole distance $D$, mass accretion rate $\dot{M}$, the radial extend of the corona $r_{\rm C}$ (the corona extends from $r_{\rm ISCO}$ to $r_{\rm C}$) 
the corona half opening angle $theta_{\rm C}$, the coronal electron temperature $\tau_{\rm C}$, the corona optical depth $\tau_{\rm C}$ 
(measured vertically above the disk to the upper and lower edge of the corona), the albedo (i.e., the reflectivity of the disk, relative to 100\% reflection), 
as well as the angle $\chi$ between the black hole axis and the celestial north pole (positive for an anti-clockwise rotation). 
Note that the reflecting disk reaches down through the corona to $r=r_{\rm ISCO}$. 
For low optical depths, the 100\% reflectivity of the disk enhances the intensity of the coronal emission substantially as photons 
scattering in the corona backwards toward the disk experience much larger energy gains than photons scattering forward  away from the disk and towards the observer. Owing to the steeply falling energy spectrum of the coronal emission, the larger energy gains translate into a substantial flux enhancement.

We fit the Stokes $I$,  $Q$, and  $U$ energy spectra for all three {\it IXPE} telescopes, the NICER Stokes $I$ energy spectrum, and the {\it NuSTAR} Stokes $I$ energy spectra of the two {\it NuSTAR} telescopes of Observations 1 and 2
({\it NuSTAR} IDs 80802313002, 80802313004). The fits include constant scaling factors to account for the flux variability between the {\it IXPE}, NICER, and {\it NuSTAR} observations.  

Overall, we find that {\tt kerrClight} does not give statistically acceptable fits, and we do not report the $\chi^2$-values here. 
Even though the fits are not perfect, they are interesting as they confront the experimental data with the predictions of the standard thin disk accretion disk model. 

\begin{figure}
\begin{center}
\includegraphics[width=0.35 \textwidth]{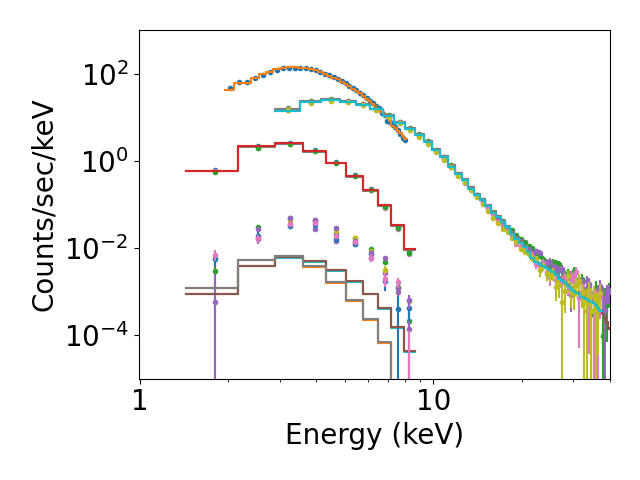}
\includegraphics[width=0.35 \textwidth]{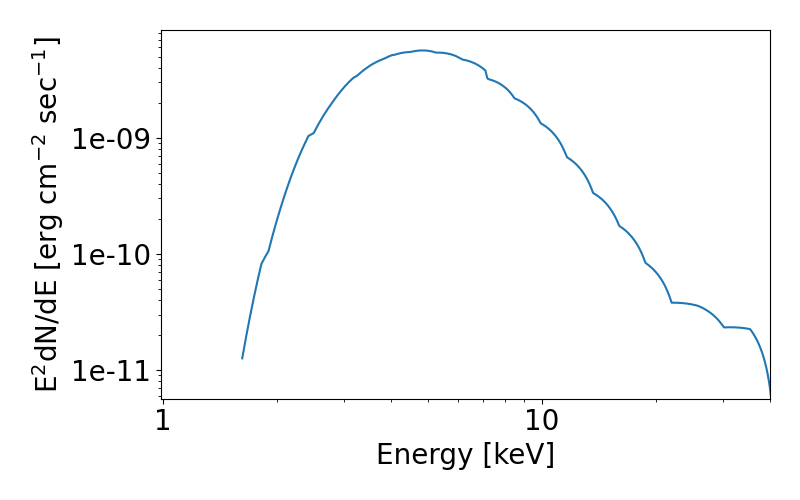}
\includegraphics[width=0.35 \textwidth]{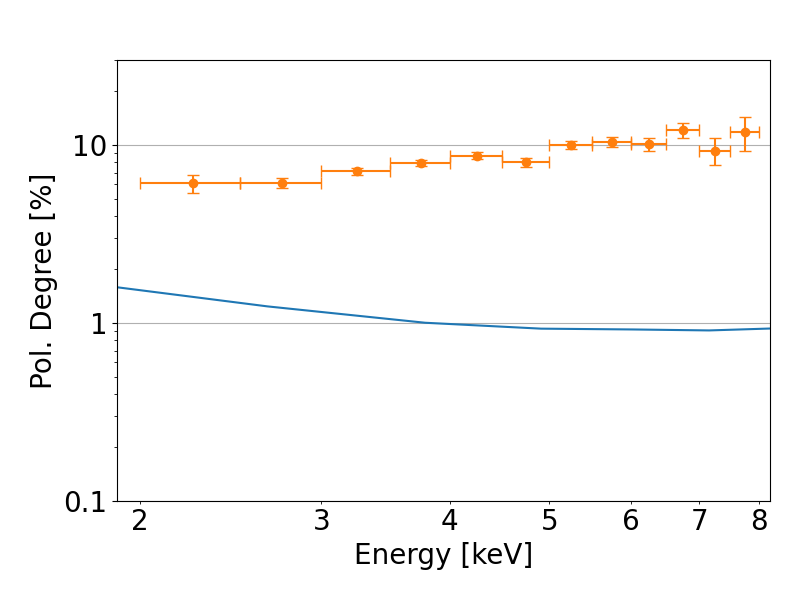}
\includegraphics[width=0.35 \textwidth]{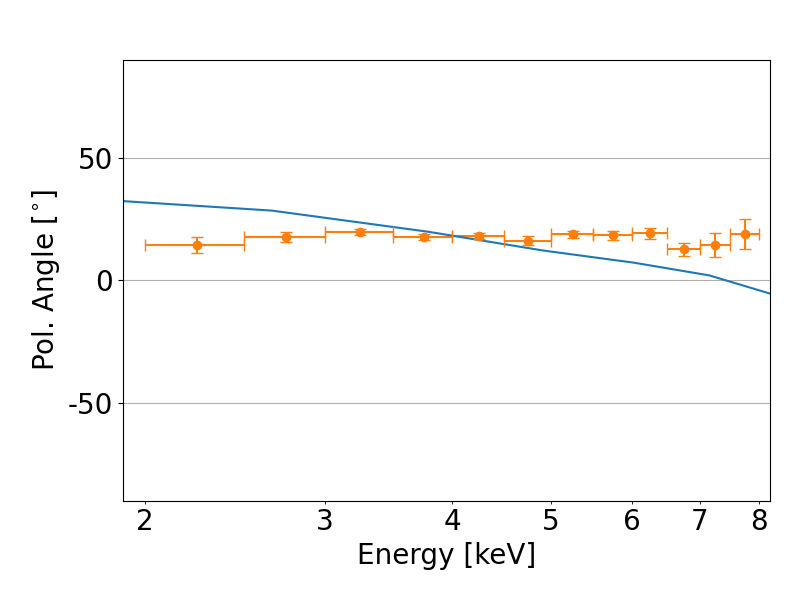}
\caption{\label{kcl1} Exemplary {\tt kerrClight} result assuming for black hole mass of 10\,$M_{\odot}$, a system distance of 11.5\,kpc,
and a 70$^{\circ}$ inclination. The panels show the observed and modeled flux and Stokes $I$, $Q$, and $U$ energy spectra (top), the model SED (upper center), PD (lower center), and PA (bottom) as a function of energy.  In the upper panel, the data points show from the top to the bottom the NICER energy spectrum, the energy spectra of the two {\it NuSTAR} telescopes for two observations, the energy spectra (Stokes $I$) of the three {\it IXPE} detectors, and Stokes $Q$ and $U$ energy spectra of the three {\it IXPE} telescopes (see Table \ref{t1} for the fit parameters). 
In the bottom panel, the PA is measured east from the celestial north with the angle increasing for a counterclockwise rotation looking towards the source. For the model, the black hole spin axis is at a PA of 52$^{\circ}$.82.   
} 
\end{center}
\end{figure}

We show two exemplary {\tt kerrClight} models. The first model (see Fig.\,\ref{kcl1} and Table\,\ref{t1}) 
assumes a fiducial  black hole mass of 10\,$M_{\odot}$, 
a distance of 11.5\,kpc \citep{2018ApJ...859...88K},
and an inclination of 70$^{\circ}$ consistent with the presence of dips but the absence of eclipses in the X-ray light curves of 4U\,1630$-$47 \citep{1998ApJ...494..747T,1998ApJ...494..753K}.
We chose a black hole spin parameter of $a\,=\,0.75$, a
corona extending from 1 to 25 $r_{\rm g}$, a corona temperature of $T_{\rm C}\,=$\,100\,keV, and a  coronal optical depth $\tau_{\rm C}$ of 0.0035 to obtain a good description of the broadband 
Spectral Energy Distribution (SED) measured with NICER,
{\it IXPE} and {\it NuSTAR}. We fit the neutral hydrogen column density $n_{\rm H}$, the mass accretion rate $\dot{M}$, and the PA $\chi$ of the black hole and accretion disk spin axis. The model predicts $\sim$1\% PDs and can thus not account for the {\it IXPE} results between 6\% and 10\%. 
The result of the modeled PDs being much lower than the observed ones is valid for a large portion of the parameter space.

\begin{figure}
\begin{center}
\includegraphics[width=0.35 \textwidth]{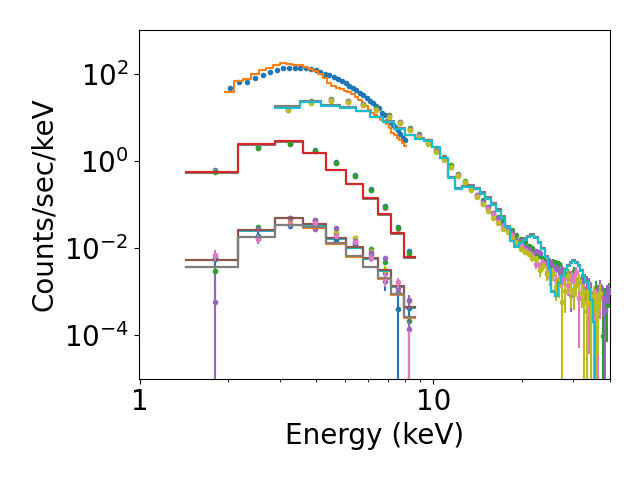}
\includegraphics[width=0.35 \textwidth]{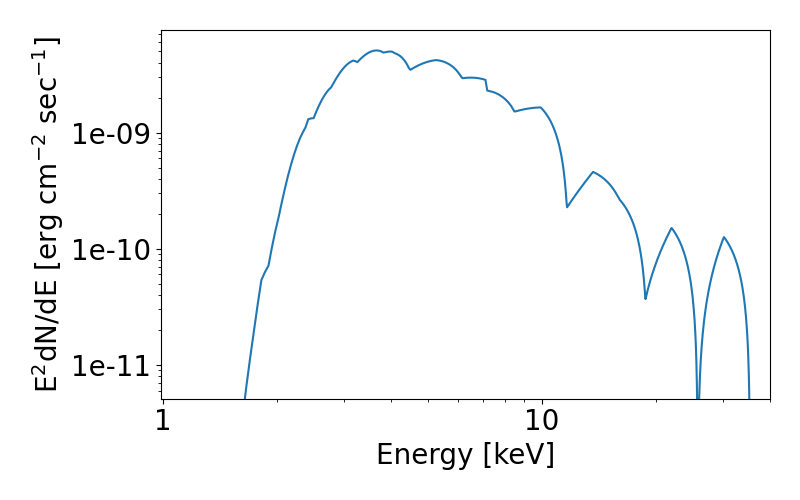}
\includegraphics[width=0.35 \textwidth]{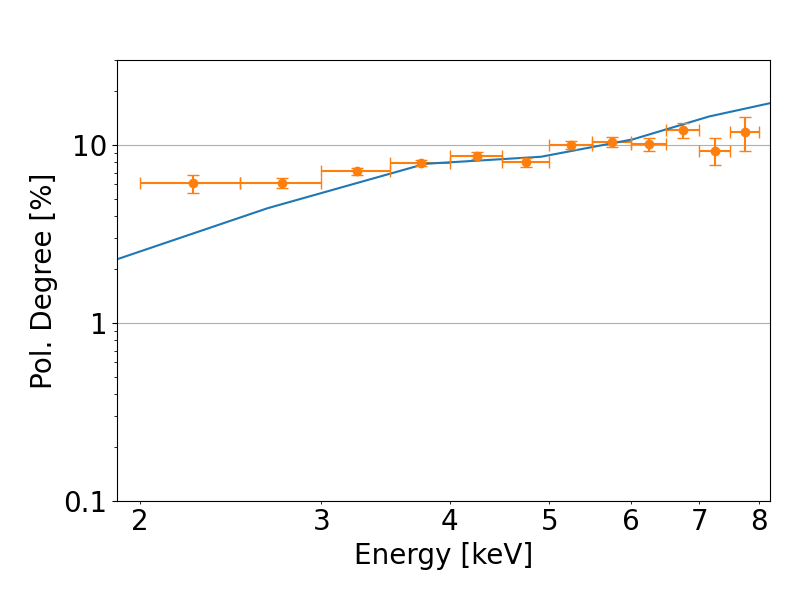}
\includegraphics[width=0.35 \textwidth]{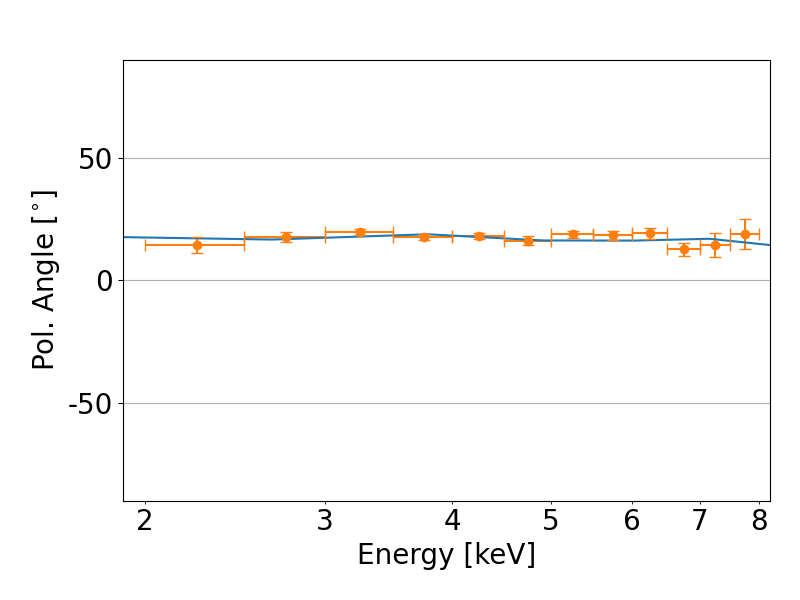}
\caption{\label{kcl2} Same as Fig.\,\ref{kcl1} but for extreme parameters that can reproduce the {\it IXPE} results (see Table \ref{t1} for the fit parameters). 
For the model, the black hole spin axis is at a PA of -11$^{\circ}$.83. Note that the model assumes a thin accretion disk without any corona; the {\tt kerrClight} optical depth parameter $\tau_{\rm C}$ is thus zero. The model has a very low disk temperature and most photons do not make it to high energies, resulting in large statistical fluctuations of the {\tt kerrClight} predictions above a few keV.}
\end{center}
\end{figure}

We forced {\tt kerrClight} to give a better description 
of the X-ray polarization results by arbitrarily 
multiplying the {\it IXPE} errors of Stokes $I$ 
and Stokes $Q$ and $U$ by factors of $10^{-2}$ ($I$) and $10^{-6}$ ($Q$ and $U$). The preferred fits combine high black hole masses of $\sim 40 M_{\odot}$ with small system distances of 
$\le 1$\,kpc. Such small distances are inconsistent 
with the distance constraint $D\,>\,(4.63\pm0.25)$\,kpc
  from the condition
that the source is positioned behind the molecular cloud MC$-$79 \citep{2018ApJ...859...88K}.
Figure\,\ref{kcl2} shows one of the best models 
(see Table\,\ref{t1} for the model parameters) that combining 
a black hole mass of 40$M_{\odot}$ with a small distance of $D\,=\,0.35$\,kpc, and a high spin of $a\,=\,0.992$. 
The high black hole mass and small distance lead to a low accretion rate and to a low temperature scale of the multi-temperature blackbody disk emission. This makes the disk spectrum relatively soft. The high spin and thus small inner disk truncation radius leads to a high fraction of high-energy photons being emitted close to the black hole, to return to the accretion disk due to 
strong gravitational lensing, and to reflect off the disk \citep[see also][]{2009ApJ...701.1175S,steiner2024ixpeled}. Owing to the Doppler boosting and de-boosting of these photons reflecting off the disk, the reflected energy spectrum is broader than that of the disk emission. Due to the steeply falling energy spectrum, the reflected emission strongly dominates over the emitted blackbody spectrum in the {\it IXPE} energy range.
Fig.\,\ref{kcl2} shows that this model can roughly generate the observed PDs and PAs as a function of energy. Note that the model explains the SED, including the high-energy power law tail, even though it assumes a disk without a corona (i.e.\ with a {\tt kerrClight} corona optical depth of $\tau_{\rm C}\,=0$). 
The high-energy power-law tail results from some photons returning multiple times to the accretion disk owing to strong gravitational lensing, and, on average, gaining energy when scattering off the disk.
The accretion disk of a rapidly spinning black hole can thus emit a power law component just like a corona. In the former case, photons experience multiple scatterings owing to the spacetime curvature close to the black hole; in the latter case, they experience multiple scatterings owing to being (temporarily) trapped inside the corona.   
Note that other models, e.g., {\tt KERRBB} \citep{2005ApJS..157..335L}, fail to predict such power law tails as they do not model the net effect of photons scattering multiple times off the accretion disk and the frequency shifts associated with Doppler boosting the photons' wave vectors into and out of the accretion disk frame for each scattering.
As a consequence of the reflected gravitationally lensed emission dominating the X-ray polarization of this model, the electric vector polarization angle is roughly aligned with the black hole spin axis. Note that most simulated photons of the low-luminosity model of Fig.\,\ref{kcl2} have $<$2\,keV energies, and the energy spectra predicted with 20$\times 10^6$ photons have noticeable statistical errors.
\begin{deluxetable}{cccc}
\tablecaption{{\tt kerrClight} parameters of Figs.\ \ref{kcl1} and \ref{kcl2}.}
\tablewidth{0pt}
\tablehead{
\colhead{Parameter} &
\colhead{Units} &
\colhead{Fig.\,\ref{kcl1}}&
\colhead{Fig.\,\ref{kcl2}}
}
\startdata
Black hole spin {\it a}&none& 0.75$^f$& 0.992$^f$\\
Black hole mass {\it M}&$M_{\odot}$&10$^f$&40$^f$\\
Inclination $i$ & $^{\circ}$ & 70$^f$& 72.20\\  
Accretion rate $\dot{M}$ & $10^{18}$\,g\,s$^{-1}$& 4.14 & 0.074\\
Distance $D$ &kpc& 11.5$^f$ &0.35$^f$\\
$\chi^a$ &$^{\circ}$& 52.82 &-11.83\\ 
Corona optical depth $\tau_{\rm C}$ &none&0.0035$^f$&0$^{b,f}$\\
Corona electron temperature $T_{\rm C}$& keV & 100$^f$ &NA\\
Corona inner radius $r_1$& $r_{\rm ISCO}$ & 1$^f$ &NA\\
Corona outer radius $r_2$& $r_{\rm ISCO}$ & 25$^f$ &NA\\
Albedo & none & 1&1\\
H I column density $n_{\rm H}$&$10^{21} {\rm cm}^{-2}$ & 12.96 &22.50$^f$\\
\enddata
\tablecomments{$^a$ Angle of disk angular momentum vector projected into the plane of the sky east from the celestial north.\newline
$^b$ $\tau_{\rm C}\,=\,0$ means that there is only a thin disk, and no corona.\newline
$^f$ Frozen parameter.}
\label{t1}
\end{deluxetable}
\section{Energetics of outflowing plasma models}
\label{outflowing}
\citet{2024ApJ...964...77R} invoke an outflowing plasma as an explanation of the high polarization degrees of 4U~1530$-$47. In this section, we discuss the energetics required for such outflows.

Assuming an outflowing photosphere emits via Bremsstrahlung,  
we can set limits on the photospheric ion and electron densities.
For electron density $n_{\rm e}$ and ion density $n_{\rm i}$ the electron-ion Bremsstrahlung volume emissivity is \citep{rybicki2008radiative}:
\begin{equation}
\epsilon_{\rm ff}\,\approx\,1.7\times 10^{-27} \,\sqrt{T}\,n_{\rm e}\,n_{\rm i}\,\rm erg\,s^{-1} \,cm^{-3}
\end{equation}
Assuming that the emission comes from a cylinder of inner and outer radii 2\,$r_{\rm g}$ and 10\,$r_{\rm g}$ 
and height $h_{\rm max}\,=10\,r_{\rm g}$ as well as 
$n_{\rm e}\,=\,n_{\rm i}$, we infer electron and ion densities exceeding  $2.3\times 10^{19}\,\rm cm^{-3}$ so that the emitted power per area
$\epsilon_{\rm ff}\,h_{\rm max}$ equals a diluted blackbody emissivity 
$I\,=\sigma_{\rm B}\,(T/f)^4\,\approx 1.8\times 10^{23}\,\rm erg\,s^{-1}\,cm^{-2}$ for $k_{\rm B}\,T\,=\,1.1$\,keV and a hardening factor of $f=1.7$.

We can use these estimates to derive limits on the kinetic luminosity of a plasma 
moving with $\beta\,=\,0.5$ times the velocity of light:
\begin{equation}
L_{\rm kin}\,=\, (\gamma-1)\,\mu_{\rm e}\,n_{\rm e}\,M_{\rm p} \,c^2\,A\,\beta\,c\,,
\end{equation}
where $\gamma\,=\,(1-\beta^2)^{-1/2}$ and 
$\mu_{\rm e}\,\approx\,1.3$ is the mean molecular weight of the ions per electron in units of the proton mass.  For a plasma dense enough to create the observed X-ray luminosity through Bremsstrahlung ($n_{\rm e}\,>\,2.3\times 10^{19}\,\rm cm^{-3}$), a kinetic luminosity exceeding $>$100 times the Eddington luminosity would be required.  We can thus exclude this scenario.

Assuming an outflowing scattering accretion disk atmosphere (or corona) gives rise to the strong polarization requires Thomson optical depths exceeding unity, and thus a column density exceeding 1$/\sigma_{\rm T}\approx 1.5\times 10^{24}\,\rm cm^{-2}$ with $\sigma_{\rm T}$ being the Thomson cross section. Using a black hole mass of 20\,$M_{\odot}$ and positing that the scattering plasma extends no more than $h_{\rm max}\,=\,10\,r_{\rm g}$ 
above the accretion disk, implies an electron density of $n_{\rm e}\,>\,(h_{\rm max}\,\sigma_{\rm T})^{-1}\approx 5 \times 10^{16}\,\rm cm^{-3}$.  The corresponding kinetic luminosity for a ion-electron plasma is 24\% the Eddington Luminosity. The scenario of an outflowing scattering medium is thus energetically challenging but not excluded.    
\\[2ex]
    
\section{Expected impact of electron and ion anisotropies on the polarization signal}
\label{anisotropies}
Particle (electron and ion) anisotropies impact the polarization of the Bremsstrahlung and scattered emission from a plasma. 
It is well known that Bremsstrahlung and Compton scattering are intimately related processes. In the framework of the Weiz\"acker-Williams formalism the properties of Bremsstrahlung emission (including the polarization properties) can be derived by treating Bremsstrahlung processes as the scattering of the virtual photons accompanying the target particle off the impinging particle \citep{1934ZPhy...88..612W,Williams:1935dka,brau2003modern}. 
The interested reader can find discussions of the polarization
of Bremsstrahlung emission in \citep{haug2004elementary} and the
polarization of scattered emission in \citep{rybicki2008radiative}.
Hard X-ray solar flares may be strongly polarized owing to polarized Bremsstrahlung emission \citep[][and references therein]{2018A&A...612A..64S}.

Although particle anisotropies of order unity can emit X-rays polarized at levels of several ten percents (see the Monte Carlo radiation transport simulations in the appendix), the effect will likely not play a noticeable role in the case of the X-ray emission from BHXRBs in the soft state. The magnitude of the anisotropies will depend on the relative time scales for generating and damping the anisotropies. The anisotropies are expected to develop together with pressure anisotropies in the sheared plasma of differentially rotating accretion disks. The rotation will lead to the amplification of the toroidal magnetic field. As particles conserve their adiabatic invariants, the pressure perpendicular to the magnetic field $P_{\perp}$ and parallel to the magnetic field $P_{\parallel}$ evolve separately as:
\begin{equation}
    \frac{d}{dt}\left(\frac{P_{\perp}}{\rho B}\right)=0,\ 
    \frac{d}{dt}\left(\frac{P_{\parallel}B^2}{\rho^3}\right)=0,
\end{equation}
where $\rho$ is the plasma density, $B$ is the magnetic field \citep{1956RSPSA.236..112C}. The shear motion increasing $B$ without changing $\rho$ will lead to $P_{\perp}>P_{\parallel}$. On the other hand, if $B$ decreases, $P_{\parallel}>P_{\perp}$ develops instead. 
We expect that such anisotropies develop on the dynamical time scale of $\sim 2\,\pi\, r_{\rm g}/c\,\approx\,6\times 10^{-4}$ s.  

The anisotropies will be damped by several processes. First, collisions will tend to isotropize the particle distribution. 
The resulting fractional pressure anisotropy is approximately given by: 
\begin{equation}
    \frac{P_{\perp}-P_{\parallel}}{P}\sim\frac{1}{\nu}\frac{1}{B}\frac{dB}{dt} \sim \frac{u}{v_{\rm th}}\frac{\lambda}{L_u}, 
\end{equation}
where $\nu$ is the collision rate, $\lambda$ is the collisional mean free path, $v_{\rm th}$ is the thermal velocity of plasma particles, $u$ is the (shear) flow velocity and $L_u$ is the length scale over which the velocity varies \citep{2016MNRAS.461.2162K}. 
Other processes that can limit the anisotropy include kinetic instabilities caused by the pressure anisotropy itself:  the fire hose instability can develop when $P_{\parallel}/P_{\perp}>(1-2/\beta_{\parallel})^{-1}$, and a mirror instability can develop in the opposite regime $P_{\perp}/P_{\parallel}>1+1/\beta_{\perp}$, where $\beta_{\parallel}=P_{\parallel}/(B^2/8\pi)$ and $\beta_{\perp}=P_{\perp}/(B^2/8\pi)$ \citep[e.g.,][]{2014PhRvL.112t5003K}. 
These instabilities lead to microscopic fluctuations that can scatter particles, leading to collisional effects that limit the pressure anisotropy at the threshold values.  
Alternative or additional mechanisms that can produce electron or ion anisotropies include particle heating or acceleration in shocks or through magnetic reconnection.

We can use the lower limits of the electron and ion densities in the photosphere or the scattering atmosphere of the 4U 1630--47 from Section \ref{outflowing} to estimate the anisotropy degrees that we can expect.  For a scattering atmosphere we derive the following constrains. 
Assuming a neutral pure hydrogen plasma, we infer an 
electron-ion collision time scale of \citep[e.g.,][]{KunzLectureNotes2023}:
\begin{equation}
\tau_{\rm ei}\,=\,\frac{3\,\sqrt{m_{\rm e}}\, (k_{\rm B}\,T_{\rm e})^{\,\,3/2}}{4\,\sqrt{2\,\pi} n_{\rm e}\, \lambda_{\rm e}\,e^4}\,<\,7.5\times 10^{-8} s,
\end{equation}
with $m_{\rm e}$ the electron mass,
$k_{\rm B}\,T_{\rm e}\,=\,3.7$\,keV$\,\approx\,5.9\times10^{-12}$\,erg, the electron Coulomb logarithm  $\lambda_{\rm e}\approx\,20$, 
and $e$ the electron charge.
The collisional time scale for ions is:
\begin{equation}
\tau_{\rm ii}\,=\,\frac{3\,\sqrt{m_{\rm i}}\, (k_{\rm B}\,T_{\rm i})^{\,\,3/2}}{4\,\sqrt{\pi} n_{\rm i}\, \lambda_{\rm i}\,e^4}\,<\,4.6\times 10^{-6} \rm s.
\end{equation}
with the ion Coulomb logarithm $\lambda_{\rm i}\approx 20$.
These times are much shorter than the dynamical time scale of 
$6\times 10^{-4}$ s. 
If the anisotropies are generated on dynamical time scales, we thus expect the electron and ion anisotropies in a scattering atmosphere to be of the order of 
$10^{-4}$ and $10^{-2}$, respectively.    
The electron and ion anisotropies in the photosphere are even smaller.   For electron density $n_{\rm e}$ and ion density $n_{\rm i}$ the electron-ion Bremsstrahlung volume emissivity is \citep{rybicki2008radiative}:
\begin{equation}
\epsilon_{\rm ff}\,\approx\,1.7\times 10^{-27} \,\sqrt{T}\,n_{\rm e}\,n_{\rm i}\,\rm erg\,s^{-1} \,cm^{-3}
\end{equation}
Assuming that the emission comes from a cylinder of inner and outer radii 2\,$r_{\rm g}$ and 10\,$r_{\rm g}$ 
and height $h_{\rm max}\,=10\,r_{\rm g}$ as well as 
$n_{\rm e}\,=\,n_{\rm i}$, we infer electron and ion densities exceeding  $2.3\times 10^{19}\,\rm cm^{-3}$ so that the emitted power per area
$\epsilon_{\rm ff}\,h_{\rm max}$ equals the diluted blackbody emissivity 
$I\,=\sigma_{\rm B}\,(T/f)^4\,\approx 1.8\times 10^{23}\,\rm erg\,s^{-1}\,cm^{-2}$ for $k_{\rm B}\,T\,=\,1.1$\,keV and a hardening factor of $f=1.7$.
For these electron and ion densities, the electron-ion and ion-ion collisional time scales are 1.6$\times 10^{-10}$~s and $10^{-8}$~s, respectively.

In summary, the intensity of the 4U\,1630$-$47 HSS emission from the region close to the $\sim$20 solar mass black hole precludes the generation of particle anisotropies of order unity that would be required to explain the strongly polarized signal from the source.

\section{Summary and Discussion}
\label{discussion}
In this paper, we discuss the impact of several mechanisms on the polarization of the X-rays from 
4U~1630$-$47. We show that a standard thin disk with or without corona can fit the flux and polarization energy spectra only when adopting  distances $<$1\,kpc, much smaller than the lower limits on the distance of 4U~1630$-$47
from \citet{2018ApJ...859...88K}. Assuming a black hole mass of 40 $M_{\odot}$ and a distance of 0.35\,kpc, we find a model with a high black hole spin of $a\,=\,0.992$ that can generate high PDs owing to the dominance of reflected gravitationally-lensed emission in the {\it IXPE}  energy band.

Our calculations show that alternative explanations of the high polarization degrees involving relativistically moving electron-ion plasmas require high mechanical luminosities. The constraints are much weaker for a scattering electron-positron plasma. 
However, we are not aware of a mechanism to generate and accelerate a suitable electron-positron plasma close to the black hole that could intercept and scatter a large fraction of the thermally emitted X-rays. 

We demonstrate that although anisotropic particles can emit strongly polarized Bremsstrahlung and Comptonized emission, the plasma in the inner portions of BHXRBs in the soft state is likely to be too dense for noticeable particle anisotropies to develop.

A promising avenue for future work are slim disk models with other geometries than those explored by \citet{2023ApJ...957....9W}. Fitting the data may require additional shadowing or reflecting features. As mentioned in the introduction, scattering tends to produce energy-independent PDs. The apparent PD increase at higher energies could be the result of dust scattering reducing the net PDs at $<$5\,keV energies. Such scenarios can possibly be tested by confronting detailed models of candidate geometries with the spectroscopic and polarimetric data.   
\section*{Appendix - Impact of particle anisotropies on the X-ray polarization signal}
\subsection*{Monte Carlo Radiation Transport Simulations}
This appendix discusses the polarization degrees and directions of the emission from plasmas with anisotropic particles. 
The study uses Monte Carlo radiation transport simulations similar to those of \citet{1978ApJ...219..705B,2011A&A...536A..93J}. 
They account for electron anisotropies but neglect ion 
anisotropies as well as polarized free-free self-absorption. 
Ion anisotropies would amplify the effect of electron anisotropies.
Self-absorption always depolarizes a signal with the emission approaching an unpolarized signal in the limit of an optically thick plasma. We first describe the Monte Carlo code and subsequently present the results.  

\subsection*{Monte Carlo Radiation Transport Simulations}
We study two configurations making use of the scattering engine described in \citep{2017ApJ...850...14B}. 
The first configuration is used to study the polarization of X-rays
emitted by pure scattering atmospheres with anisotropic electrons. The photons are emitted at the bottom of the atmosphere extending
from $z=-5\,l_{\rm sc}$ to $z=0$ with $l_{\rm sc}$ being the scattering mean free path. 
We set the polarization of the emitted 
photons to 0 to clearly see the effect of the scatterings on the photon polarization.

Photons scatter off electrons drawn from a 
Maxwell-Boltzmann distribution of temperature $T$. 
The code uses four wave-vectors $k^{\mu}=(E,E\vec{n})$ to keep track of the energy 
$E$ and the direction $\vec{n}$ of a photon. 
The code keeps track of the photon's linear polarization with the help of a parameter tracking the PD, 
and the polarization vector $f^{\mu}=(0,\vec{f})$ with $|\vec{f}|=1$ encoding 
the electric field polarization direction 
\citep{2018grav.book.....M}.
The Compton scattering is effected by Lorentz transforming $k^{\mu}$ and $f^{\mu}$ into the 
scattering electron's rest frame. 
After drawing a random direction of the scattered 
photon, we construct the Stokes vector of the incoming photon referenced to the plane 
spanned by the wave vectors of the 
incoming and outgoing photon. 
The Stokes vector of the outgoing photon is 
calculated by multiplying the Stokes vector 
of the incoming photon with Fano's 
fully relativistic scattering matrix \citep{1957RvMP...29...74F,1961RvMP...33....8M,2017ApJ...850...14B}.
A rejection algorithm uses the Stokes-$I$ parameter of the scattered photon to account for the 
energy and scattering angle dependence of the 
Klein-Nishina cross section. 
The scattering changes the photon energy in 
the electron rest frame according to 
Compton's equation.
The Stokes vector is subsequently used to 
infer the PD and polarization direction 
$\vec{f}$ of the scattered photon. 
In the last step, the wave and polarization vectors 
are transformed back into the plasma frame.
We verified the code's performance by reproducing Chandrasekhar's results for a test run in which we switched off all relativistic effects 
\citep[change of electron energy, Klein-Nishima cross section, scattering probability as a function of the angle between electron velocity and photon wave vector, see]
[]{2017ApJ...850...14B}. The Comptonization code was furthermore cross-checked in the deep
Klein-Nishina regime against the MONK code 
(W. Zhang, private communication, 2022).

The second configuration is used to study the impact of the polarized 
Bremsstrahlung emission, photon absorption, and Compton scattering
on the polarization of the emergent emission. 
We simulate a 5-absorption-length-deep atmosphere 
at (electron) temperature $T$. 
The relative importance of emission and scattering is 
parameterized by the ratio $r_{\rm sc/a}$ of the absorption to scattering cross sections, 
and we use the parameters $n$ and $m$ to 
characterize the electron anisotropy.
Bremsstrahlung photons are emitted uniformly throughout the atmosphere. 
The PD and polarization angle are generated by making use of the relativistic cross sections $\sigma_{II}$ and $\sigma_{III}$ for the emission of photons polarized parallel and perpendicular 
to the plane defined by the electron and photon velocity vectors derived by \citet{1953PhRv...90.1030G}
in the first Born approximation. 
Note that Equations (4.2) and (4.3) of \citet{1953PhRv...90.1030G} 
are correct up to a typo (multiplication instead of 
subtraction at the beginning of the last line of Equation (4.3)) 
that Gluckstern acknowledged in a private communication mentioned 
in \citep{1978ApJ...219..705B}.
The reproductions of these equations by \citet{1978ApJ...219..705B}
include errors as do those of \citet{2016MNRAS.461.2162K}.
The latter authors give the Bremsstrahlung 
cross sections in a convenient form, but 
their Equation (16) for $L$ 
includes a factor of 2 that should be dropped.
In our code, photons propagate until they are absorbed, 
scatter, or escape the atmosphere. The Compton scatterings 
are simulated as explained above.

Photons escaping the atmosphere are sorted into 6 bins in the cosine 
of the inclination of the observer $\mu_{\rm obs}$ and in 5 bins in energy, and the Stokes parameters within each bin are summed. Owing to the symmetry of the 
plane parallel atmosphere, Stokes-$U$ vanishes. We denote electric 
field polarizations perpendicular to the atmosphere as 
positive polarization (PD=$Q/I>$0), and polarizations parallel 
to the atmosphere as negative polarization (PD=$Q/I<$0).

\subsection*{Results: Polarization of Scattered Emission}
In the first step, we use the code to demonstrate how much electron anisotropies modify the PDs compared to the classical results for a pure electron scattering atmosphere derived by Chandrasekhar \citep{Chandra:60}. 
Fitting the energy spectrum of 4U\,1630--47 during the {\it IXPE} HSS observations with the diluted multi-temperature blackbody model  {\tt ezdisk} \citep{2005ApJ...618..832Z}, gives a maximum blackbody temperature $T_{\rm BB}$ of 1.4\,keV \citep{2024ApJ...964...77R}. 
Adopting the radial temperature dependence of the  {\tt ezdisk}  model, this gives an emission weighted blackbody 
temperature of 1.1\,keV when averaging from 2\,$r_{\rm g}$ to  10\,$r_{\rm g}$. 
In the following, we assume a plasma with  electron and ion temperatures $T_{\rm e}$ and $T_{\rm i}$ of 3.7\,keV, as  Bremsstrahlung of a plasma at these temperatures generates an 
energy spectrum with a $\nu F_{\nu}$ high-energy cutoff similar to the one of a 1.1\,keV blackbody.
The simulations focus on a small local region of the accretion disk (or corona) modeled as a plane-parallel slab lying in the $x$-$y$-plane with the surface normal pointing along the $z$-direction. 
For a BHXRB, the $z$ axis would be parallel to the 
angular momentum vector of the inner accretion disk and presumably also parallel to the black hole angular momentum vector. 
The global effects from transporting the emission through the curved spacetime of a Kerr black hole, and from reflecting the emission off the accretion disk are not considered here. Adding up the emission from different regions of the accretion disk will lower the net polarization degree. 

\begin{figure}
\begin{center}
\includegraphics[width=0.5 \textwidth]{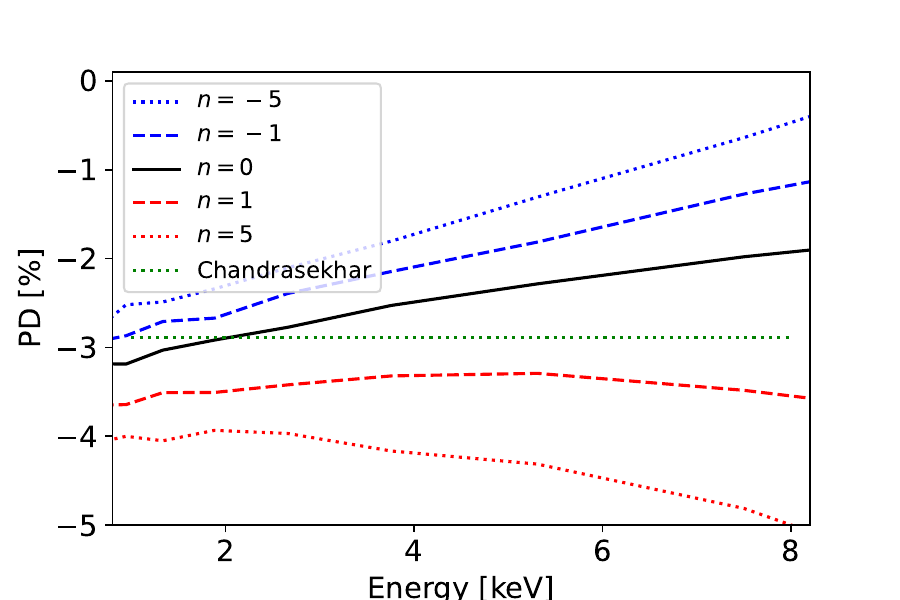}
\end{center}
\caption{\label{f:chandra} PDs produced by a pure electron scattering atmosphere at temperature $k_{\rm B}T\,=\,$3.7\,keV for an observer at inclination $i=65^{\circ}$.
Positive PDs show a polarization perpendicular to the emitting plasma slab, and  negative PDs show a polarization parallel 
to the emitting plasma slab.
The lines show the results for $n=-5$ (blue dotted line),
$n=-1$ (blue dashed line), $n=0$ (black solid line), 
$n=1$ (red dashed line) and $n=5$ 
(red dotted line).
The green dotted line shows 
Chandrasekhar's result for reference.
}
\end{figure}

We consider first the case in which the electrons move preferentially up and down in the emitting slab with a probability distribution:
\begin{equation}
p(\mu)\propto (\mu^2)^n \label{e1}
\end{equation}
with $n>0$ and $\mu$ being the angle between the electron velocity and the $z$-axis. We parameterize the opposite behavior, electrons moving preferentially in the $x$-$y$-plane of the emitting slab, as:
\begin{equation}
p(\mu)\propto (1-\mu^2)^{|n|} \label{e2}
\end{equation}
for $n<0$. 
Throughout this paper, we adopt the convention that positive (negative) PDs represent electric vector polarization directions perpendicular (parallel) 
to the emitting plasma slabs.
Figure \ref{f:chandra} shows the results for the likely 
65$^{\circ}$ inclination of 4U\,1630--47. 
Preferred electron motion perpendicular (parallel) to the slab with $n>0$ ($n<0$) increases (decreases) 
the polarization parallel to the slab.
Interestingly, the polarization degrees depend on energy owing to the correlation of the photon energy and the mean number of scatterings. 
We conclude that pronounced anisotropies can impact the PDs significantly.  
\begin{figure}
\begin{center}
\includegraphics[width=0.45\textwidth]{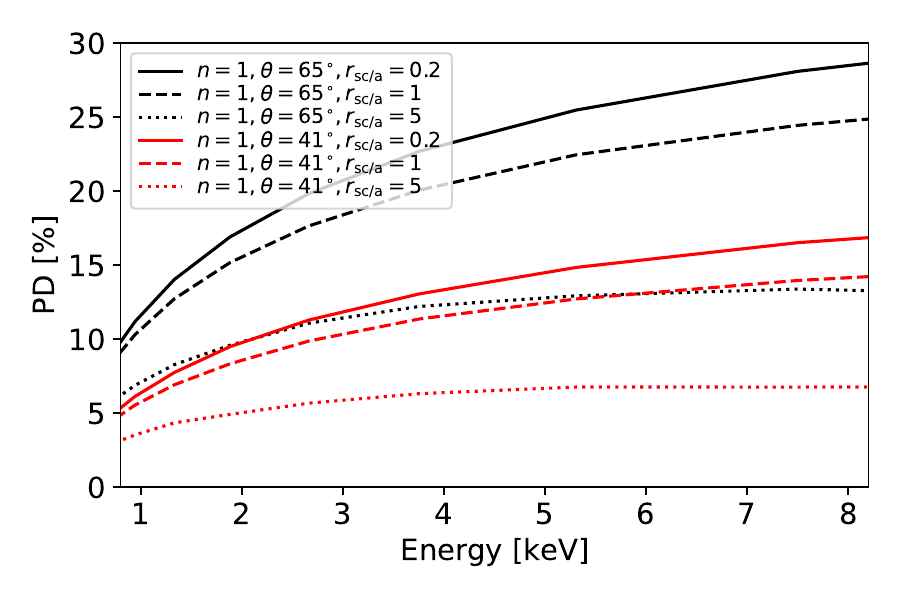}
\hspace*{-0.5cm}
\includegraphics[width=0.475\textwidth]{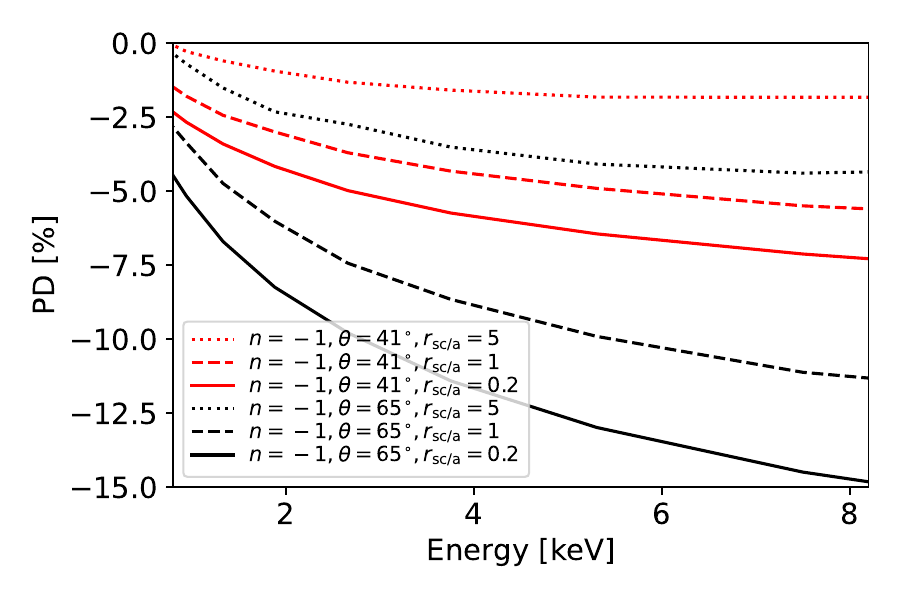}
\end{center}

\caption{\label{f:updown} 
PDs produced by electrons moving preferentially 
perpendicular to the atmosphere ($n=1$, upper panel) and
parallel to the atmosphere ($n=-1$, lower panel) 
for observers at $i=65^{\circ}$ (black lines)
and $i=41^{\circ}$ (red lines).
For each color, the different lines show the results for different
scattering to absorption cross section ratios:
$r_{\rm sc/a}=0.2$ (solid lines),
$1$ (dashed lines), and
$5$ (dotted lines).}
\end{figure}

\subsection*{Results: Polarization of Bremsstrahlung Emission}
As mentioned above, the Bremsstrahlung emission from anisotropic particles is polarized. For photon energies less than $\sim$1/10th of the electron energy, the Bremsstrahlung is polarized perpendicular to the electron direction as a consequence of small-angle electron scatterings. Conversely, for photon energies exceeding $\sim$1/10th of the electron energy, the Bremsstrahlung is polarized parallel to the electron direction as a consequence of large-angle electron scatterings.  The polarization degree reaches 100\% for photons  with energies close to the energy of the emitting electron.  As mentioned above, we assume a plasma temperature of 3.7\,keV.
Figure \ref{f:updown}  shows the polarization energy spectra for the same electron anisotropies as above,  for various ratios $r_{\rm s/a}$ of grey scattering-to-absorption cross sections.  
For $n=+1$ (electrons preferentially moving up and down), the scenario gives polarizations perpendicular to the surface normal with the 8\,keV polarization degrees reaching $\sim$30\% for the most optimistic scenario that we simulated. For $n=-1$, the polarization direction is parallel to the disk and reaches $\sim$15\% at 8 keV for the most optimistic simulated scenario.   

\begin{figure}[t!]
\begin{center}
\includegraphics[width=0.5\textwidth]{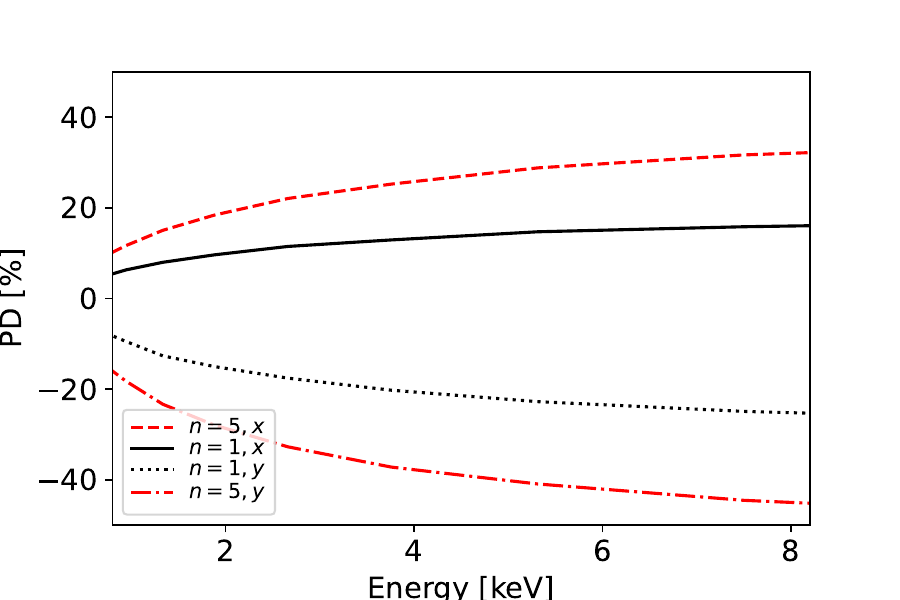}
\end{center}
\caption{\label{f:m+1} 
Polarization from electron with direction symmetric around the $x$-axis for $n=+1$ (black lines) and $n=+5$ (red lines)
as seen for an $i=65^{\circ}$ observer 
viewing the  atmosphere along the direction 
of the symmetry $x$-axis (dashed and solid lines) and 
perpendicular to it in the plane of the atmosphere 
($y$-axis, dotted and dash-dotted lines)
for $r_{\rm sc/a}=1$.
}
\end{figure}

The shearing motion of the differentially rotating accretion flow is expected to produce a toroidal magnetic field in the accretion flow. For adiabatic changes of the magnetic field, the Electrons gyrating around the field lines can develop anisotropies regarding the pitch angle of the electrons relative to the magnetic field lines. We studied such a scenario with electrons described by Equation (\ref{e1}), but with $\mu$ being the pitch angle cosine relative to the $x$-axis, assumed to be parallel to the magnetic field direction. Figure \ref{f:m+1} shows the PDs for this scenario. The results are somewhat more complicated as the PDs and PAs depend on the position of the observer relative to the $x$-axis. The net polarization would result from the superposition of the polarization from different regions of the disk. Global simulations would be required to show which polarization wins, taking into account strong gravitational lensing and gravitational and Doppler frequency shifts.   
\\[4ex]
{\it Acknowledgments:}
The authors acknowledge very helpful comments by the referee C.\,Done, including drawing our attention to the possibility of dust scattering creating the apparent rise of the polarization degree with energy. The authors thank the {\it IXPE} team for many fruitful discussions of the {\it IPXE} results.  H.K.\ thanks Banafsheh Beheshtipour for the original code of the Compton scattering engine. The Washington University authors thank the McDonnell Center for the Space Sciences for financial and logistic support. H.K., N.R.C., K.H.\ and S.C.\ acknowledge NASA support through the grants 80NSSC20K0329, 80NSSC21K1817, 80NSSC22K1291, 80NSSC22K1883, 80NSSC23K1041, 80NSSC24K0205, and 80NSSC24K1178.  
A.C.\ and Y.Y.\ acknowledge support from NSF grants DMS-2235457 and AST-2308111. A.C.\ also acknowledges NASA support from grant 80NSSC21K2027. Y.Y.\ also acknowledges support by the Multimessenger Plasma Physics Center (MPPC), NSF grant PHY-2206608. The Washington University authors acknowledge support from the McDonnell Center for the Space Sciences. N.R.C. acknowledges support by the John Templeton Foundation.
\bibliography{apjl}
\end{document}